\documentclass[twocolumn,showkeys,preprintnumbers,floatfix,amsmath,a4paper,nofootinbib,amssymb]{revtex4}
\newcommand{\beginsupplement}{%
        \setcounter{table}{0}
        \renewcommand{\thetable}{S\arabic{table}}%
        \setcounter{figure}{0}
        \renewcommand{\thefigure}{S\arabic{figure}}%
     }
\usepackage{graphicx}

\usepackage[utf8x]{inputenc}
\usepackage[russian,english]{babel}

\begin{document}
\selectlanguage{english}
\title{Gender Bias in Sharenting: Both Men and Women Mention Sons More Often Than Daughters on Social Media}

\author{Elizaveta Sivak$^1$ and Ivan Smirnov$^1$}

\thanks{ibsmirnov@hse.ru}

\affiliation{
$^1$ Institute of Education; National Research University Higher School of Economics, Myasnitskaya ul., 20, Moscow 101000, Russia}

\begin{abstract}
Gender inequality starts before birth. Parents tend to prefer boys over girls, which is manifested in reproductive behavior, marital life, and parents' pastimes and investments in their children. While social media and sharing information about children (so-called ``sharenting'') have become an integral part of parenthood, it is not well-known if and how gender preference shapes online behavior of users. In this paper, we investigate public mentions of daughters and sons on social media. We use data from a popular social networking site on public posts from 635,665 users. We find that both men and women mention sons more often than daughters in their posts. We also find that posts featuring sons get more ``likes'' on average. Our results indicate that girls are underrepresented in parents' digital narratives about their children. This gender imbalance may send a message that girls are less important than boys, or that they deserve less attention, thus reinforcing gender inequality.
\end{abstract}

\maketitle

Gender inequality starts even before birth. Across the world, would-be parents tend to prefer their first (or the only) child to be a boy rather than a girl or to have more sons than daughters
\cite{tian2018gender,hank2000gender,rossi2015gender,abeykoon1995sex,bongaarts2013implementation,dahl2003preconception,dahl2006preconception,raley2006sons}. This results in millions of ``missing girls'' at birth due to sex-selective abortions \cite{duthe2012high,miller2001female,world2012world}. Gender preference continues to manifest throughout childhood. In some countries couples pursue sons by having additional children at the cost of larger family size and underinvestment in daughters \cite{zaidi2016pursuit,altindag2016preference}. Sons have advantages in nutrition \cite{barcellos2014child}, vaccination rates \cite{borooah2004gender} and spending on health care \cite{song1999search,ganatra1994male}. Fathers \cite{baker2016boy,harris1991fathers,aldous1998fathering} (and in some cases both parents \cite{lundberg2005division}) spend more time with sons than with daughters. Fathers more often marry and stay married in families with sons \cite{dahl2008demand,blau2017there}, although evidence for this is mixed \cite{lundberg2007child,diekmann2004parents}. At this point the reader will not be surprised that parents also report more happiness in families with sons \cite{kohler2005partner}.

Despite the extensive literature on gender preference, no study to date has examined whether the use of social media by parents is gender biased. As social media becomes an integral part of parents' life, it might be important to understand if and how gender preference is manifested in this environment. One common practice that has recently become a widespread trend is ``sharenting'' \cite{blum2017sharenting,brosch2016child}, or parents' habitual use of social media to communicate a lot of detailed information about their children\footnote{Sharenting, as cited in Collins Dictionary}. In this paper, we investigate gender preference in sharenting drawing on data about 62 millions public posts on a popular social networking site. 

\begin{figure}
\centering
\includegraphics[width=1.0\linewidth]{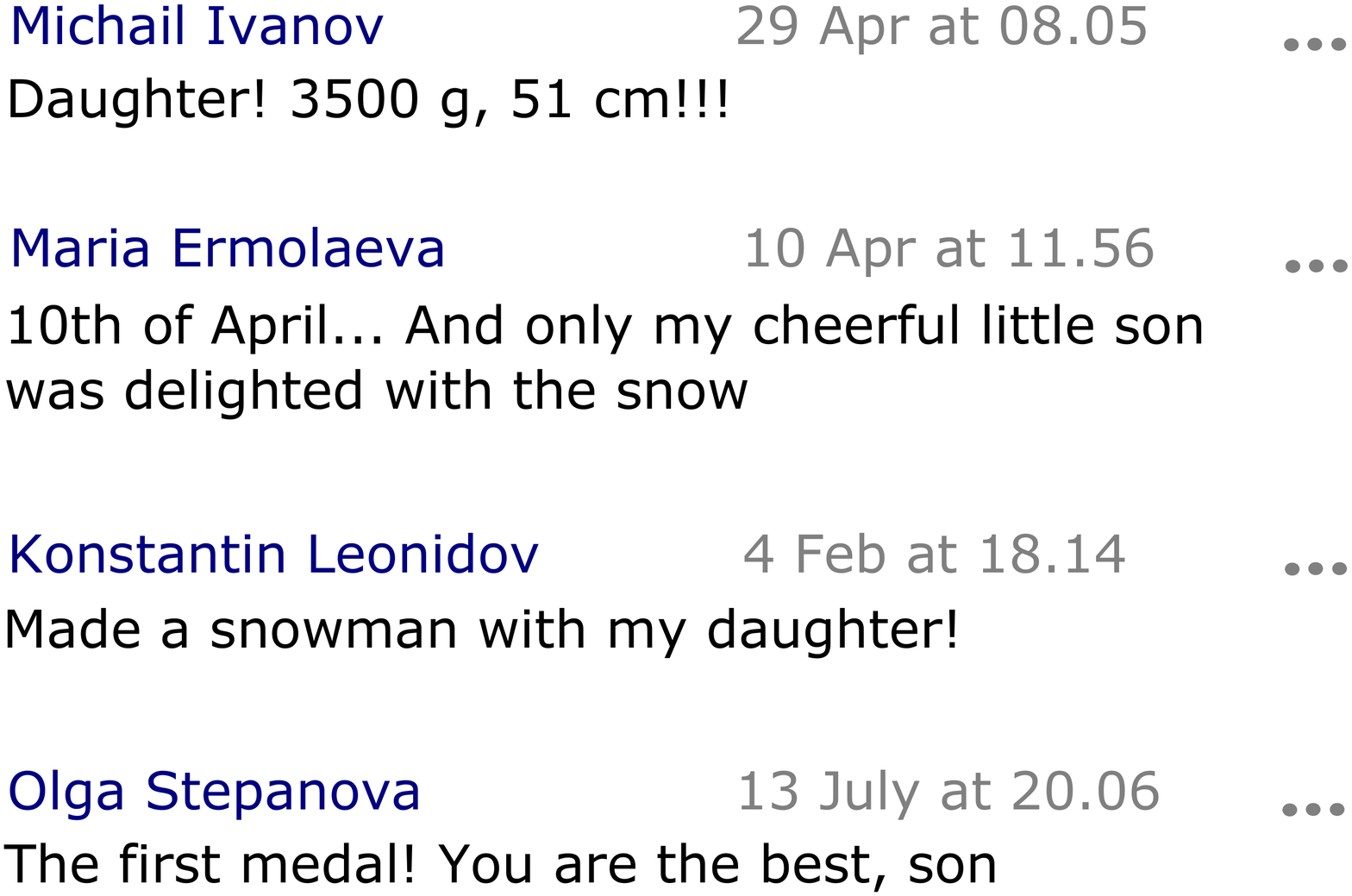}
\caption{Selected examples of posts with mentions of children. All names and dates have been changed.}
\label{fig:Examples}
\end{figure}

We obtained data from VK\footnote{http://vk.com}, a Russian analogue of Facebook and the largest social networking site in Europe. VK provides an application programming interface (API) that allows the systematic download of publicly available information. We used the VK API to collect public posts that were made in 2016 by 635,665 users from Saint Petersburg (fourth largest European city), aged 18-50 years (see SI Text for details on our sample and data collection). We then identified posts with mentions of children by looking at posts that contained the words ``daughter'' and ``son'' along with their different forms, e.g. ``dochenka'' (daughterling) or ``soooooooon'' (see Methods for details). Common topics for such posts included celebrations of different achievements and important events (e.g. births and birthdays or starting and finishing school), expression of love, affection, and pride, and reports on spending time with the children (see Fig. \ref{fig:Examples} for illustrative examples and SI Text for more information about common topics).

We computed the proportion of female and male users from each cohort who mentioned sons or daughters in their posts at least once along with the average number of mentions of children for these users. In our analysis, we used various definitions for ``mentions of children'' to ensure that the results were not influenced by a specific choice of words (see Methods). We also collected information about the number of ``likes'' that posts featuring children obtained, on average. We used these data to investigate whether the social network environment might reinforce gender bias by rewarding posts featuring children of one gender more than those of another.

\section*{Results}
\begin{figure}
\centering
\includegraphics[width=1.0\linewidth]{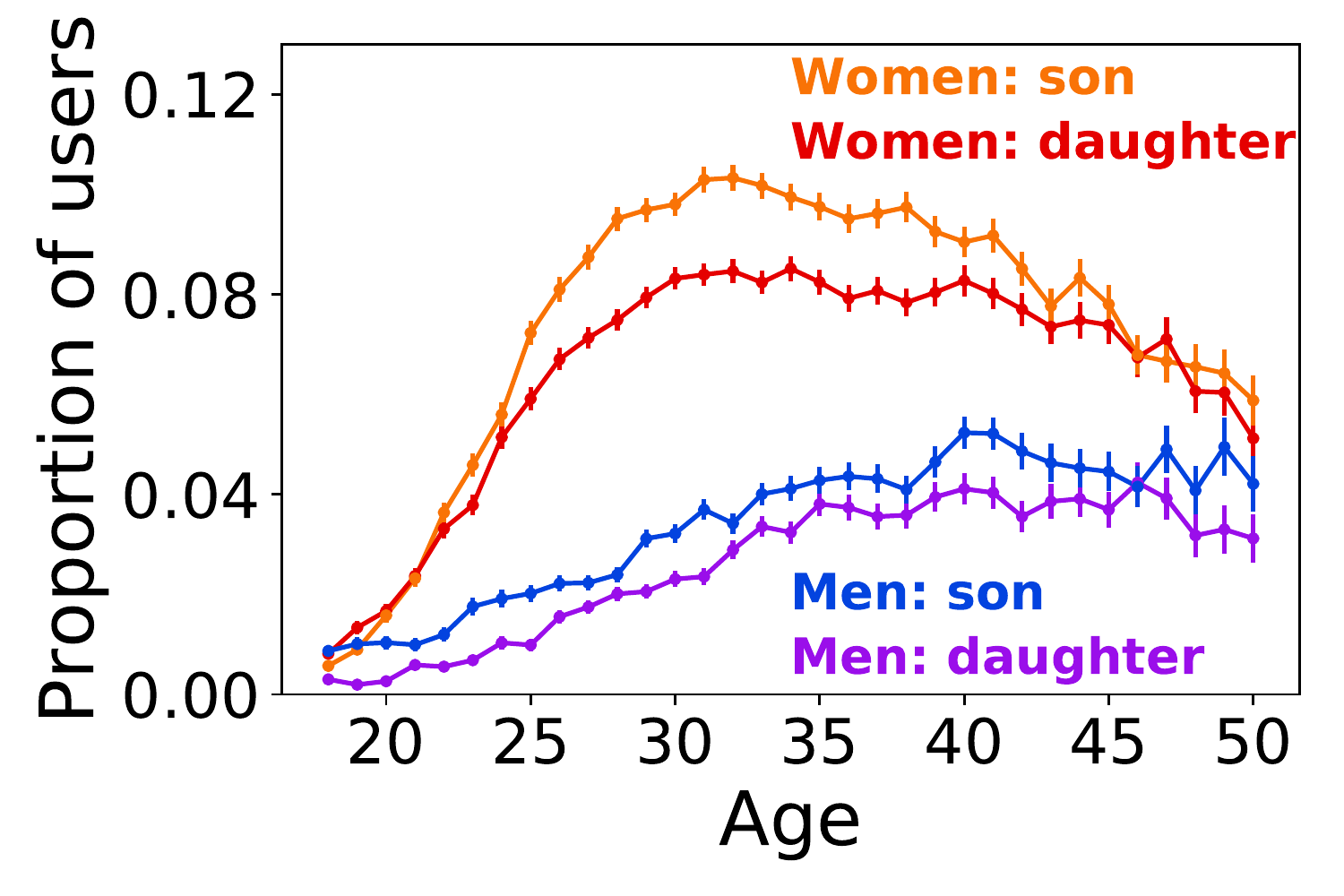}
\caption{The proportion of users who mentioned children in their public posts at least once in 2016. Sons are mentioned by a larger proportion of both men and women. Vertical bars are standard errors.}
\label{fig:Main}
\end{figure}

Fig. \ref{fig:Main} shows the proportion of users who mentioned children in their public posts at least once in 2016. The proportion of women increases sharply until 31-32 years old and then gradually falls. The peak matches the average age of women at first childbirth, which is 30 years in Saint Petersburg \cite{interfax20176highest}. The proportion of men who mention children is significantly lower and steadily increases with age.

In almost all cohorts of users, sons are mentioned by a larger proportion of both men and women. This difference cannot be explained by the sex ratio at birth alone (1.06 in Russia) and thus indicates gender preference in sharing information about children. Those users who mention children at least once also write slightly more posts about sons. There are 2.3 posts about sons per woman and 2.1 posts about daughters per woman ($p = 0.0001$). Men write 1.7 posts about sons and 1.5 posts about daughters on average ($p = 0.05$). As a result of these two tendencies, there are more posts about sons than about daughters on the social network. The exact estimate of the gap in the number of posts depends on the set of words that are chosen as synonyms for the words ``son'' and ``daughter'' (see SI Text for detailed analysis). From our most conservative estimate, women write 15\% more posts about sons than about daughters, and men write 43\% more posts about sons.

We also found that posts featuring sons are more rewarded, that is they get more ``likes'', than those featuring daughters. Average numbers of ``likes'' are presented in Table \ref{tbl:Likes}. Here three patterns can be distinguished. First, women ``like'' posts more often than men. Second, there is a gender homophily in ``likes'', i.e. women prefer posts written by women and men prefer those written by men. Third, both women and men more often ``like'' posts which mention sons.

\begin{table}
\centering
\caption{The average number of likes per post. All daughters/sons differences are statistically significant with $p < 10^{-4}$.}
\begin{tabular}{lrrr}
Written by women & \multicolumn{2}{l}{Average number of likes} \\
& by women & by men \\
\hline
featuring daughters & 6.7 & 1.1 \\
featuring sons & 10.7 & 1.8 \\
\hline
\\
\end{tabular}

\centering
\begin{tabular}{lrrr}
Written by men & \multicolumn{2}{l}{Average number of likes} \\
& by women & by men \\
\hline
featuring daughters & 5.3 & 2.6 \\
featuring sons &  6.7 & 3.7 \\
\hline
\end{tabular}
\label{tbl:Likes}
\end{table}

\section*{Discussion}
Studies of gender preference in parental practices usually have to rely on self-reports, e.g. reports about time spent with children \cite{baker2016boy,harris1991fathers,aldous1998fathering,lundberg2005division}. Self-report studies have some benefits, but their results are affected by various biases including social desirability bias or recall bias. Mentions in posts are directly observable and present a clear and simple metric, which can be used on easily accessible data to measure parents' gender bias. We used this metric on a large dataset of public posts of more than six hundred thousand users and found that both men and women exhibited son preference on the social networking site: sons were mentioned significantly more often than daughters. This result is remarkably stable, and holds true across age cohorts, different measures, and sets of words. We also found that writing posts in which sons are mentioned is more rewarded: these posts get around 1.5 times more likes than stories featuring daughters.

Son preference in traditional societies and developing countries is a well-known phenomenon. Our results confirm that son preference is also prevalent in countries not immediately associated with gender disparity\footnote{Russia is above average in the ranking of countries by gender parity \cite{wef2016global}.}. 

Gender preference in ``sharenting'' may seem quite harmless in comparison with such layers of gender disparity as sex-selective abortions or underinvestment in girls. However, son bias online may affect girls as they may feel underappreciated and less visible. It may also have broader effects on gender parity. Even moderate bias might accumulate given the widespread popularity of social media. Son preference in ``likes'' can additionally amplify the bias, acting as social media's built-in positive feedback loop. Millions of users are exposed to a gender biased news feed on an everyday basis and, without even noticing, get the reaffirmation that it is normal to pay more attention to sons.

Previous studies have shown that children's books are dominated by male central characters \cite{mccabe2011gender,hamilton2006gender}. In textbooks, females get fewer lines of text, fewer named characters, and fewer mentions than men \cite{blumberg2008invisible}. Additionally, in movies there are on average twice as many male characters as female ones in front of the camera \cite{smith2015inequality}. While female coverage on Wikipedia compares favorably with some other lists of notable people \cite{wagner2015s}, there are still 4 times more articles about men than women \cite{wiki2018gender}. Gender imbalance in public posts may send yet another message that girls are less important and interesting than boys and deserve less attention, thus presenting an invisible obstacle to gender equality.

\section*{Methods}
\subsection*{Counting mentions of children}
We used the API of VK to download all public posts of users from Saint Petersburg that were made in 2016. We then computed vector representations of Russian words by training a fastText \cite{bojanowski2016enriching} model on the collected corpus. We used this model to identify words similar to ``son'' and ``daughter'', namely the closest words in the vector space measured by cosine distance. We manually excluded unrelated words. For instance, both the words ``son'' and ``granddaughter'' are unsurprisingly semantically close to the word ``daughter'' according to the model. However, these are not synonyms to the word ``daughter'' and we do not treat them as mentions of daughters. After exclusion of unrelated words we obtained a list of the 30 closest synonyms to the word ``daughter'' and the 30 closest synonyms to the word ``son''. The posts that included at least one of these words were considered as posts mentioning children.
The use of word embeddings trained on the VK corpus allowed us to take into account words or their forms that cannot be found in dictionaries but which are used by the users of the social network, e.g. ``sooon'' instead of ``son''. We performed an additional analysis to make sure that our results were not driven by a particular choice of words (see SI Text). 
We also removed potentially fake accounts and filtered posts that were not made by users themselves (see SI Text for details on data preprocessing) and then computed the proportion of users who mentioned children at least once in their posts, and the average number of such mentions per user.

\section*{Acknowledgements}
Support from the Basic Research Program of the National Research University Higher School of Economics is gratefully acknowledged.

\bibliographystyle{unsrt}
\bibliography{genderbias}

\newpage
\beginsupplement
\section*{Supplementary Information}
\subsection*{Sample and data preprocessing}
VK is the largest European social networking site, with more than 100 million active users. It was launched in September 2006 in Russia and provides functionality similar to Facebook. According to VK's Terms of Service: ``Publishing any content on his / her own personal page, including personal information, the User understands and accepts that this information may be available to other Internet users taking into account the architecture and functionality of the Site''. VK provides an application programming interface (API) that enables downloading of information systematically from the site. In particular, it is possible to download user profiles from particular educational institutions and within selected age ranges. For each user, it is possible to obtain a list of their public posts. VK's support team confirmed to us that the data downloaded via their API may be used for research purposes.

One limitation of VK is that its API returns no more than 1000 users for any request. To collect data on users from Saint Petersburg on a larger scale we created a list of all high schools in Saint Petersburg and then accessed IDs of users from each age cohort (from 18 to 50 years) who graduated from each of these schools. As each of these requests returns less than 1000 users, we were able to collect information about all users who indicated their high school on VK. Note that not all users in the sample currently live in Saint Petersburg, and not everyone on VK indicated their (former) high school in their profile. However, this approach allowed us to collect a large sample of VK users in a systematic way. Another advantage of our approach is that it provides an opportunity to effectively remove fake profiles. To achieve this we did not include in the final sample the users who had no friends on the social networking site from their high school. The data were collected as part of the Digital Trace project and the data collection procedure was approved by the Institutional Review Board of the National Research University Higher School of Economics.

We made sure that only posts with authentic content were included in the analysis. We excluded reposts and posts with exactly the same content made by multiple users. We also did not include posts containing URLs to accounts for potential automatic posting and advertisements by websites or VK applications (e.g. invitations to visit a web site or to beat the score in a game). Not all the posts in the resulting sample are necessarily about users' own children. Some posts include mentions of children of friends or relatives, or jokes, etc. By our estimate, the proportion of such posts is around 9\%, and their removal does not affect the observed son bias (see SI Topic analysis).

\subsection*{Dictionary analysis}
The exact estimate of son bias might depend on the selection of words that are counted as mentions of children. However, we found that the observed preference for sons holds true irrespective of the choice of words. To show this, we selected the N most frequent synonyms and forms of the words ``daughter'' and ``son'' from our corpus. We then used these sets of words to compute the proportion of users who mentioned children at least once (Fig. S1a), as well as the total number of posts with mentions of children (Fig. S1b). The son bias holds true for all $N = 1,...,40$. Any changes for larger N are negligible. The list of the most frequent words along with the number of occurrences of each word is presented in Table S1.

\begin{figure}
\centering
\includegraphics[width=1.0\linewidth]{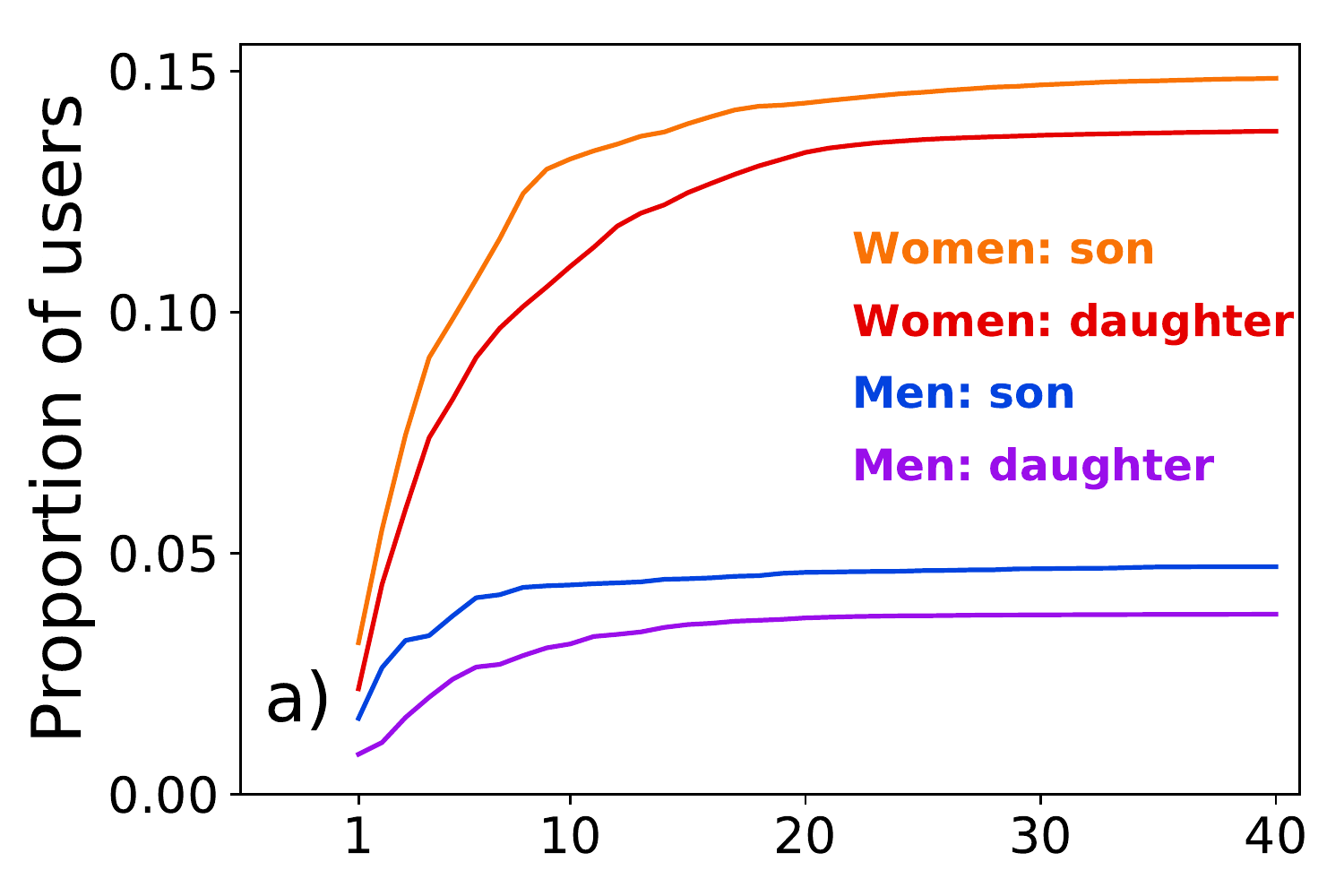}
\includegraphics[width=1.0\linewidth]{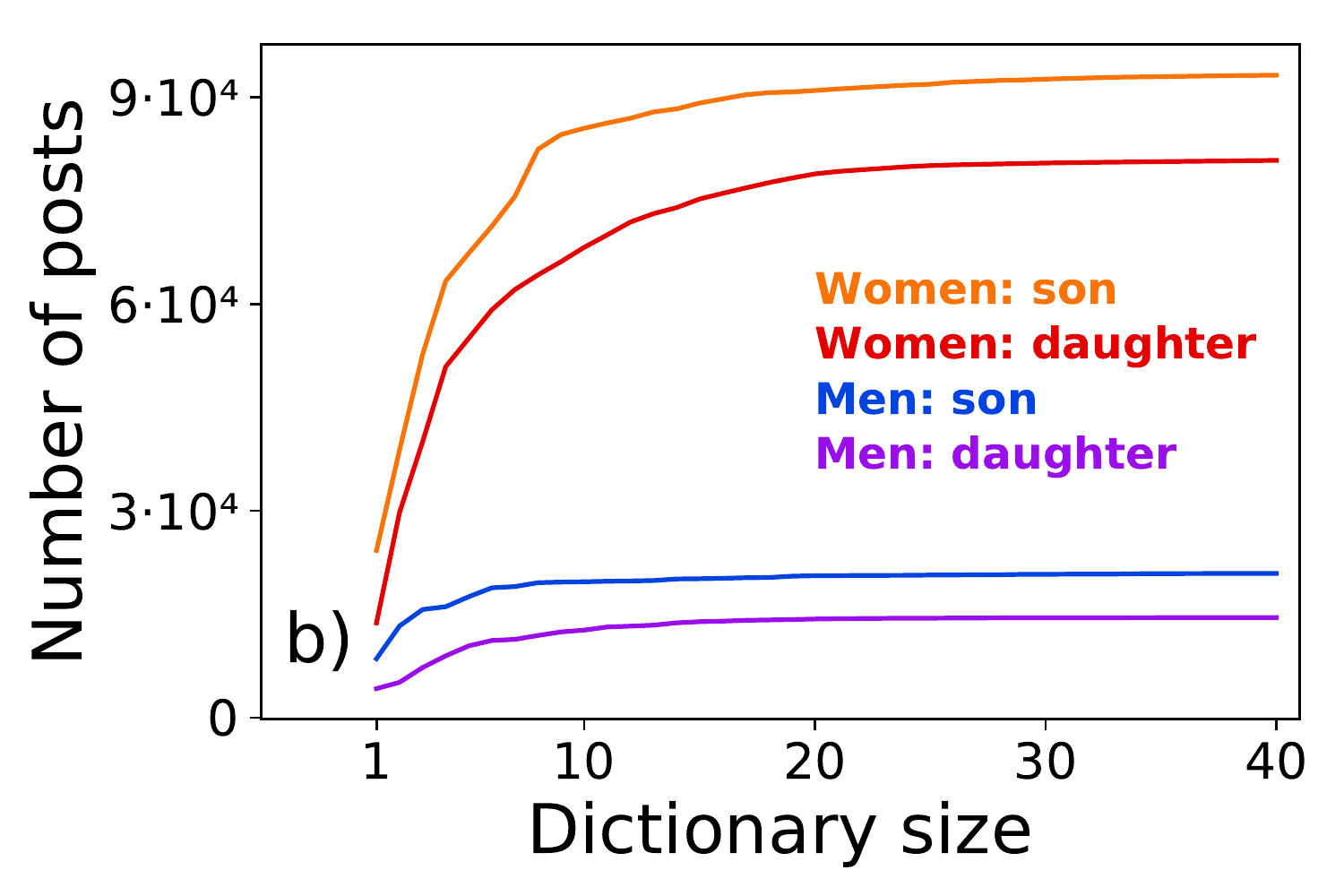}
\caption{Proportion of users who mention children at least once (a) and the total number of posts with mentions of children (b), as a function of the number of words included in the analysis.}
\label{fig:Dictionary}
\end{figure}

\subsection*{Topic analysis}
We identified the main topics of posts with mentions of children by analyzing a sample of posts from one age cohort (30 years old). We coded all the men's posts (879 posts) from this cohort and randomly selected 20\% (1521) of the women's posts. At the first stage of coding, for each post we wrote down the category which most fully grasped the post's content. At the second stage we collapsed similar categories into broader ones. 

Only 9\% of the posts are not related to the users' own children (8\% among women's posts and 18\% among men's). These posts include mentions of other people's kids as well as jokes, news, and stories about pets. After filtering out the irrelevant posts, the son bias for women was unchanged, and for men it remained significant: women wrote 15\% more posts about sons than about daughters and men wrote 34\% more posts about sons in this age group.

Among relevant posts, the most common topics were reports of spending time with children (27\% of posts), expressions of positive feelings, mostly love, affection, or pride (26\%), and celebrations of births and birthdays (19\%; see Fig. S2 for examples). These three categories accounted for 72\% percent of all posts about user's own children. Note that the distribution of topics most likely depends on the age of a child, and might be different for other cohorts.

\begin{figure}
\centering
\includegraphics[width=1.0\linewidth]{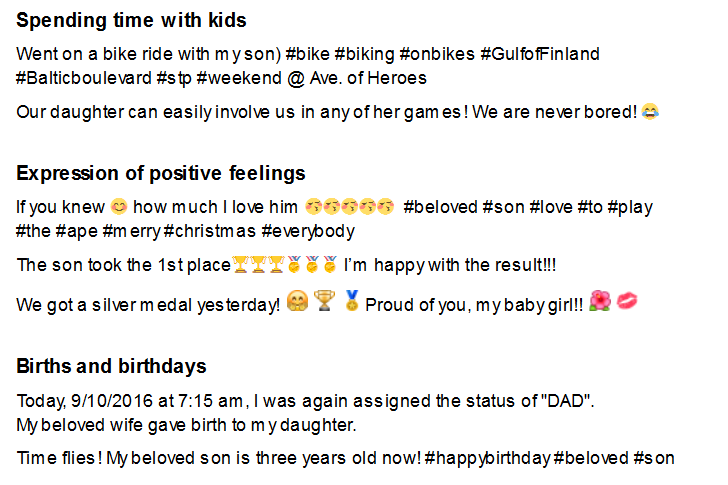}
\caption{Examples of posts featuring children. Authors of the posts provided consent to the use of their posts in this publication. All posts were originally written in Russian and were translated verbatim.}
\label{fig:Examples2}
\end{figure}

\begin{table}
\centering
\caption{The list of the most frequent words along with the number of occurrences of each word.}
\selectlanguage{russian}
\begin{tabular}{lrllr}
сын & 27133 &~& дочь & 16010 \\
сына & 18218 &~& доченька & 10477 \\
сынок & 11564 &~& дочка & 9245 \\
сыночек & 7443 &~& доча & 7350 \\
сыном & 5308 &~& дочери & 5415 \\
сыну & 4953 &~& дочки & 4493 \\
сыночка & 4123 &~& доченьки & 2622 \\
сынуля & 4086 &~& дочкой & 2597 \\
сыночку & 2091 &~& дочку & 2413 \\
сынули & 877 &~& дочей &2214 \\
сыночком & 793 &~& дочке & 2148 \\
сынулей & 698 &~& доченьке & 1861 \\
сынулька & 685 &~& дочи &  1229 \\
сыне & 660 &~& дочерью & 1193 \\
сыночки & 609 &~& дочурка & 1067 \\
сынуле & 606 &~& доченьку & 871 \\
сынишка & 593 &~& доче & 819 \\
сынулю & 323 &~& доченькой & 755 \\
сынов & 293 &~& дочурки & 659 \\
сынка & 250 &~& дочу & 629 \\
сынульки & 216 &~& дочурке & 375 \\
сынишки & 198 &~& дочурку & 268 \\
сынульке & 188 &~& дочуркой & 241 \\
сыночков & 182 &~& дочура & 120 \\
сынишке & 154 &~& дочечка & 117 \\
сынулик & 151 &~& дочуля & 94 \\
сынишку & 147 &~& дочурок & 83 \\
сыночке & 136 &~& дочкин & 73 \\
сынки & 119 &~& дочечки & 58 \\
сынулечка & 108 &~& дочуня & 58 \\
сынишкой & 103 &~& дочкины & 49 \\
сынульку & 98 &~& доченек & 47 \\
сынулькой & 96 &~& дочин & 39 \\
сынку & 84 &~& дочуры & 37 \\
сынком & 59 &~& дочушка & 36 \\
сыночкам & 56 &~& доченькам & 35 \\
сынишек & 52 &~& дочурочка & 35 \\
сынуличка & 40 &~& дочкина & 34 \\
сынкам & 40 &~& дочик & 30 \\
сыночками & 38 &~~~~~~~~~& дочкою & 30
 
\end{tabular}
\end{table}

\end{document}